\documentclass[prl,twocolumn,superscriptaddress,showpacs]{revtex4}
\usepackage{graphicx}
\usepackage{amsmath}
\usepackage{amssymb}

\newcommand{\llo}{L1$_0$}

\begin{document}

\title{Spin-orbit strength driven crossover between intrinsic and extrinsic
mechanisms of the anomalous Hall effect in epitaxial \llo\ FePd and FePt}

\author{K. M. Seemann}
\altaffiliation{Present address: Institut f\"{u}r Festk\"{o}rperforschung
(IFF-9), FZ J\"{u}lich GmbH, D-52425 J\"{u}lich, Germany}
\email{k.seemann@fz-juelich.de}
\affiliation{School of Physics and Astronomy, University of Leeds, Leeds,
LS2 9JT, United Kingdom}

\author{Y. Mokrousov}
\affiliation{Institut f\"{u}r Festk\"{o}rperforschung and Institute for
Advanced Simulation, Forschungszentrum J\"{u}lich, D-52425 
J\"{u}lich, Germany}

\author{A. Aziz}
\affiliation{Department of Materials Science and Metallurgy, University of
Cambridge, Pembroke Street, Cambridge, CB2 3QZ, United Kingdom}

\author{J. Miguel}
\affiliation{Freie Universit\"{a}t Berlin, Institut f\"{u}r Experimentalphysik,
Arnimallee 14, D-14195 Berlin, Germany}

\author{F. Kronast}
\affiliation{Helmholtz-Zentrum Berlin f\"{u}r Materialien und Energie,
Albert-Einstein-Strasse 15, Berlin, Germany}

\author{W. Kuch}
\affiliation{Freie Universit\"{a}t Berlin, Institut f\"{u}r Experimentalphysik,
Arnimallee 14, D-14195 Berlin, Germany}

\author{M. G. Blamire}
\affiliation{Department of Materials Science and Metallurgy, University of
Cambridge, Pembroke Street, Cambridge, CB2 3QZ, United Kingdom}

\author{A. T. Hindmarch}
\affiliation{School of Physics and Astronomy, University of Leeds, Leeds,
LS2 9JT, United Kingdom}

\author{B. J. Hickey}
\affiliation{School of Physics and Astronomy, University of Leeds, Leeds,
LS2 9JT, United Kingdom}

\author{I. Souza}
\affiliation{Department of Physics, University of California, Berkeley, CA
94720, USA}

\author{C. H. Marrows}
\email{c.h.marrows@leeds.ac.uk}
\affiliation{School of Physics and Astronomy, University of Leeds, Leeds,
LS2 9JT, United Kingdom}

\date{\today}

\begin{abstract}
We determine the composition of intrinsic as well as extrinsic contributions 
to the anomalous Hall effect (AHE) in the isoelectronic \llo~FePd and FePt
alloys. We show that the AHE signal in our 30 nm thick epitaxially deposited 
films of FePd is mainly due to extrinsic side-jump, while in the epitaxial 
FePt films of the same thickness and degree of order the intrinsic contribution 
is dominating over the extrinsic mechanisms of the AHE. We relate this crossover 
to the difference in spin-orbit strength of Pt and Pd atoms and suggest that 
this phenomenon can be used for tuning the origins of the 
AHE in complex alloys.
\end{abstract}


%71.70.Ej    Spin-orbit coupling, Zeeman and Stark splitting, Jahn-Teller effect
%72.15.-v    Electronic conduction in metals and alloys
%75.50.Bb    Fe and its alloys
%75.60.Ch    Domain walls and domain structure
%75.25.+z    Spin arrangements in magnetically ordered materials
%            (including neutron and spin-polarized electron studies,
%            synchrotron-source X-ray scattering, etc.)
%76.30.He    Platinum and palladium group (4d and 5d) ions and impurities (Zr-Ag and Hf-Au)

\maketitle

The anomalous Hall effect (AHE) in a ferromagnet as discovered by 
E. H. Hall in 1881~\cite{hall1881}, can have its physical origin in 
bandstructure effects,~i.e.~be of intrinsic nature~\cite{karplus1954}, 
or be driven by the scattering of the conduction electrons at magnetic 
impurities, which is called extrinsic AHE. The two basic extrinsic 
effects leading to AHE are the skew~\cite{smit1958} and 
side-jump~\cite{berger1970} scattering, and both are commonly understood 
to be a consequence of the spin-orbit interaction (SOI) acting on a 
conduction band electron. While many discoveries in the field of AHE 
are being made in both, experiments~\cite{branford2009,takahashi2009,tian2009} 
but also computationally~\cite{yao2004,roman2009}, a joint effort is 
essential to explain the results of experiments with modern AHE theory. 
This becomes obvious when fitting the measured anomalous Hall resistivity 
$\rho_{\rm H}$ to the sum of linear and quadratic terms in diagonal 
resistivity $\rho_{\rm 0}$:
\begin{equation}\label{eq1}
\rho_{\rm H} = \Phi_{\mathrm{Sk}}\rho_0 + \left(\kappa^{\rm sj} -
\frac{e^2}{8\pi^3\hbar}\int_{\rm BZ}\Omega(\mathbf{k})d^3k
\right)\rho_0^2.
\end{equation}
In this relation the linear term stands for skew-scattering~\cite{smit1958}, 
and the intrinsic contribution, given by an integral of the Berry curvature
$\Omega(\mathbf{k})$ over pristine crystal's occupied electronic 
bands~\cite{berry1984,nagaosa2006}, provides the Karplus-Luttinger 
part \cite{karplus1954} of the anomalous Hall conductivity. Side-jump
conductivity $\kappa^{\rm sj}$ accumulates all the other extrinsic
contributions to $\rho_{\rm H}$, quadratic in 
$\rho_{\rm 0}$~\cite{berger1970,nagaosa2006}.
Among the three contributions in Eq.~(\ref{eq1}), the intrinsic part can 
be calculated reasonably well from first-principles, 
while only model expressions for the extrinsic AHE exist up to now~\cite{nagaosa2006}.
The ability to experimentally adjust the interplay between the 
different contributions of the AHE is of utter importance for the 
deeper understanding of this phenomenon, also with respect to developing 
a microscopic theory of extrinsic scattering mechanisms based on material's 
electronic structure. This may also be particularly valuable in prospect of 
novel applications at room temperature and above.

The epitaxially ordered ferromagnets \llo -FePd and -FePt are ideal to
evaluate the degree to which both extrinsic and intrinsic effects
interplay in the experimentally observed AHE as they offer the possibility
to controllably alter the strength of the SOI by exchanging isoelectronic
atomic species Pd and Pt. Their well-defined crystal structure makes
them amenable to theoretical analysis. Other physical phenomena such as
domain wall resistance \cite{marrows2005} of \llo -ordered FePt and FePd
have received in-depth investigation before
\cite{ravelosona1999,marrows2004,seemann2007}. Whilst there have been
measurements of the AHE in these materials
\cite{viret2000,yu2000,mihai2008,moritz2008}, a direct comparison
between \llo -ordered FePt and FePd with respect to AHE has not been
carried out. In this Letter, the underlying physical effect of
spin-orbit scattering in \llo -ordered FePt and FePd is investigated,
and we show in a combined experimental and theoretical study that
whilst the AHE in FePd arises predominantly from extrinsic side-jump
scattering, the much stronger SOI in FePt gives rise to a large intrinsic
Berry phase AHE dominating over the side-jump contribution to the measured
AHE signal.

Our experiments were performed on epitaxial thin films of \llo -ordered
FePd and FePt, which were deposited onto polished single crystalline
MgO(100) substrates by dc magnetron co-sputtering as described in
Ref.~[\onlinecite{seemann2007}]. The epitaxial quality, i.e. the
degree of chemical long range ordering $S_{\rm Order}$, was determined
by X-ray diffraction, whilst the nominal film thicknesses were confirmed 
by X-ray reflectometry. We have deliberately selected films of comparable 
thickness $t$ of $31\pm1$~nm (FePd) and $34\pm1$~nm (FePt), and degree of 
chemical ordering $S_{\rm Order}=0.8\pm0.1$ to be able to compare the 
magnetotransport properties as closely as possible.
\begin{table}
\begin{ruledtabular}
\caption{
Ordinary and anomalous Hall coefficients $R_\mathrm{0}$ and $R_\mathrm{H}$
for the \llo -ordered FePt and FePd epilayers at $T=50$ and 270~K.}
\label{tabD}
\begin{tabular}{lllll}
     &  $R_\mathrm{0}$ (50 K) & $R_\mathrm{0}$ (270 K) & $R_\mathrm{H}$ (50 K) 
     & $R_\mathrm{H}$ (270 K) \\
 & $\mathrm{n\Omega cm}/\mathrm{T}$ & $\mathrm{n\Omega cm}/\mathrm{T}$ 
 &  $\mathrm{\mu\Omega cm}/\mathrm{T}$ & $\mathrm{\mu\Omega cm}/\mathrm{T}$ \\
\hline
  FePd &  $-28 \pm 3$ & $-16 \pm 2$ & $0.01 \pm 0.01$ & $0.14 \pm 0.05$ \\
  FePt &  $-3 \pm 1$ & $-2 \pm 1$ & $0.12 \pm 0.05$ & $0.75 \pm 0.10$ \\
\end{tabular}
\end{ruledtabular}
\end{table}

\begin{figure}
   \includegraphics[width=3.2cm]{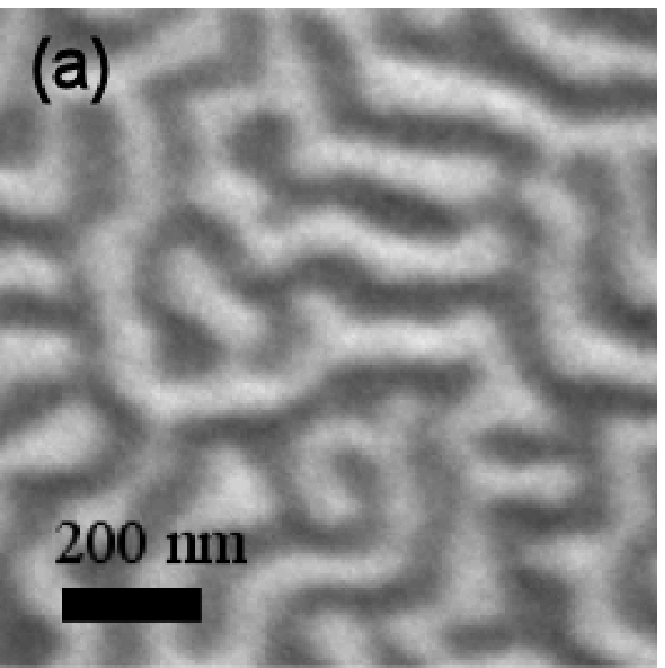}
   \includegraphics[width=3.2cm]{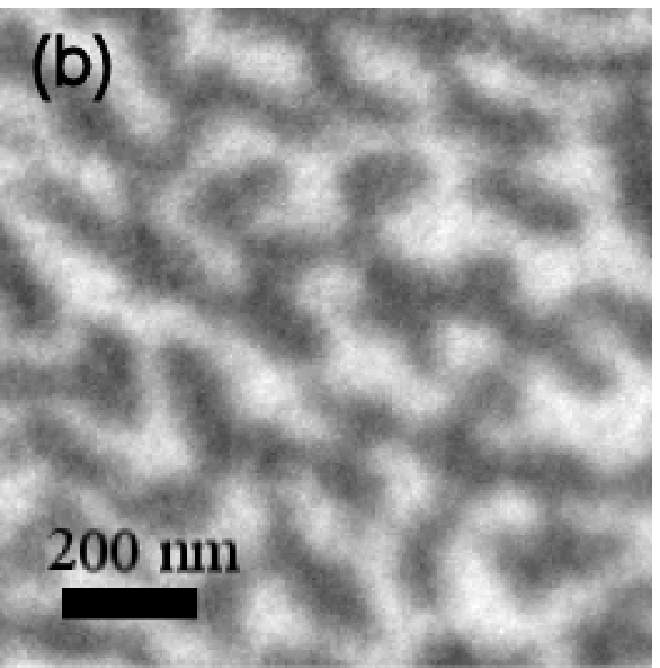}\\
\caption{XMCD-PEEM images of 1 $\mu$m $\times$ 1 $\mu$m area of the
room temperature demagnetized domain state of (a) \llo -ordered FePd
and (b) FePt films in zero magnetic field. Dark and bright areas depict
magnetic domains of oppositely orientated magnetization perpendicular
to the film plane. \label{domains}}
\end{figure}

The highly anisotropic \llo -crystal lattice and strong SOI from
the 4$d$ and 5$d$ species give rise to a very strong magnetocrystalline
anisotropy with a uniaxial easy axis normal to the film plane. This
yields a dense labyrinth domain pattern after {\it ac} demagnetization.
We imaged this zero-field domain state in both films by means of
X-ray magnetic circular dichroism-photoelectron emission microscopy
(XMCD-PEEM) \cite{stoehr1993} at the UE49-PGM-a microfocus beamline
of the BESSY II facility. We used the XMCD effect at the Fe $L_3$-absorption
edge at 707 eV to image the magnetic domains of the FePt and FePd thin
films, and the resulting images are shown in Fig. \ref{domains}. The
images are represented as greyscale coded absorption asymmetry for
opposite helicities of the circularly polarized X-rays,
$A_{\rm PEEM}=\frac{I_+-I_-}{I_++I_-}$. For constant absolute value
of the magnetic moments, $A_{\rm PEEM}$ is proportional to the
projection of the local magnetization on the direction of incidence
of the light.

Both domain patterns show the characteristic maze structure 
obtained at zero-field for the \llo -ordered Fe-alloys 
investigated~\cite{yu2000,seemann2007}. From the $1\times1$ 
$\mathrm{\mu m}^2$ area PEEM images we determine the average 
domain width of
$\overline{D}_{\mathrm{PEEM}}=69\pm20$ nm for FePd and
$\overline{D}_{\mathrm{PEEM}}=125\pm30$ nm for FePt, typical values for
films of this thickness and degree of crystallographic order\cite{marrows2004,seemann2007,seemann2008}. High crystallographic 
long range order parameter of $S_{\rm Order}\approx0.8$ determined 
from X-ray diffractometry together with the experimental 
verification by XMCD-PEEM of the demagnetized domain state 
simulated from measured micromagnetic parameters serve as evidence 
for the \llo -phase of our samples. 
\begin{figure}
  \includegraphics[width=7.0cm]{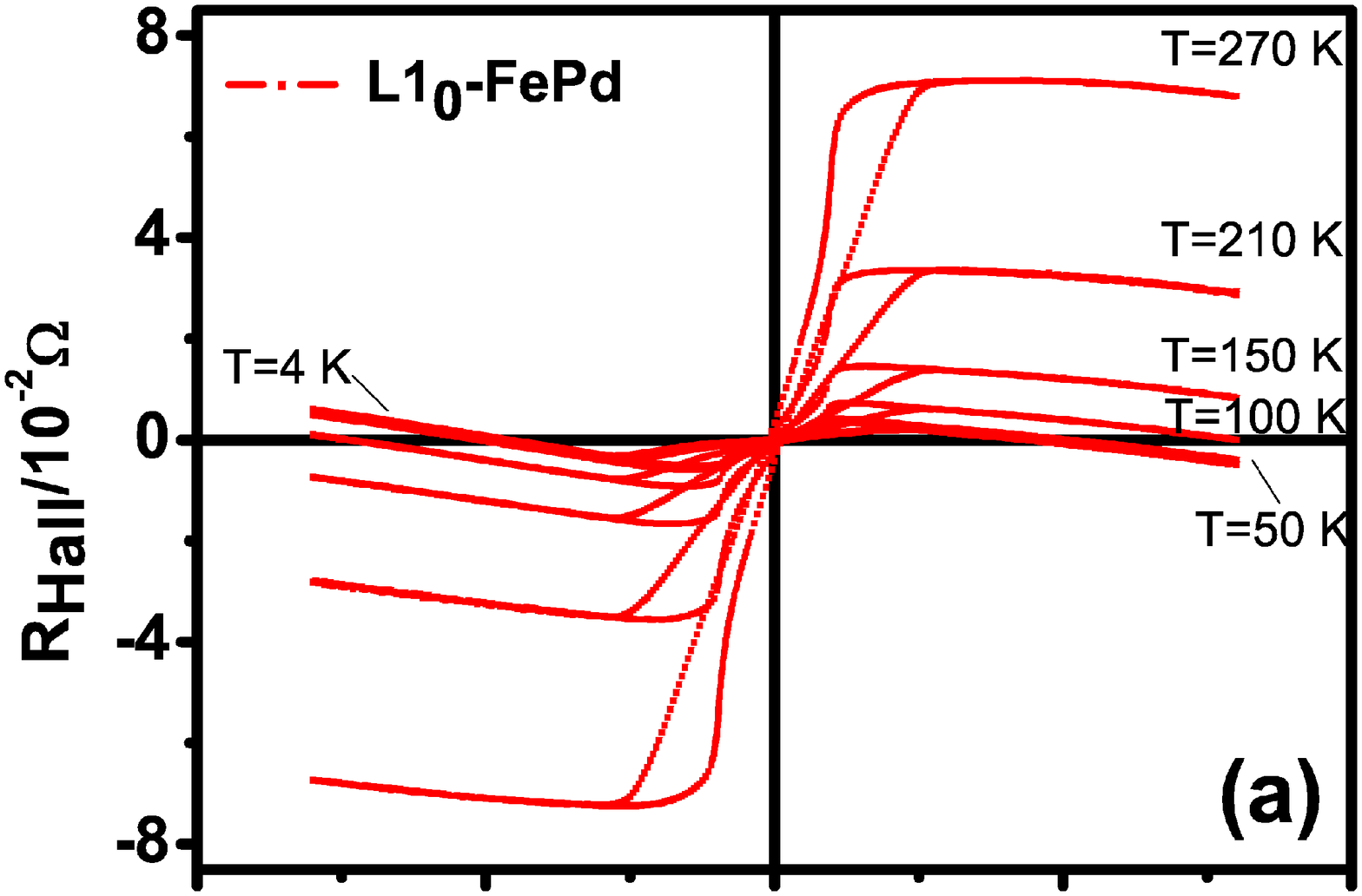}
 \includegraphics[width=7.0cm]{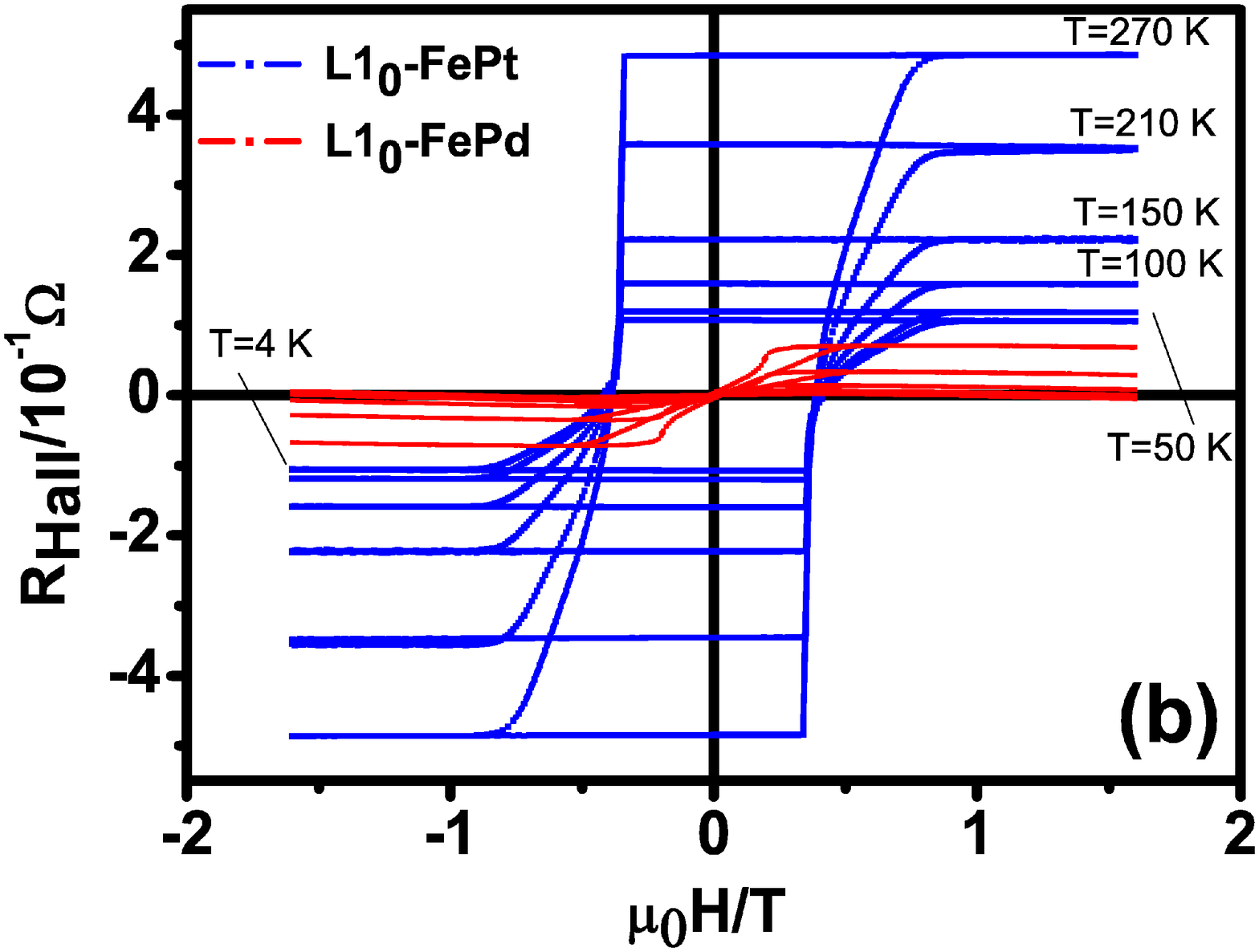}
  \caption{(Color online) Measured Hall resistance $R_{\mathrm{Hall}}$
for various temperatures ranging between $T=4$ K to $270$ K for (a)
\llo -FePd and (b) \llo -FePt epilayers. The measured AHE signal is,
for all temperatures, much larger in \llo -FePt as compared to
\llo -FePd, as shown in (b).}
  \label{AHEmeasured}
\end{figure}

The typical hysteretic property of the Hall resistance 
$R_{\mathrm{Hall}}$ is obtained from 4-probe {\it dc} Hall 
measurements and shown in Fig.~\ref{AHEmeasured}(a) and (b) 
for both \llo -ordered FePd and FePt. Describing the total 
Hall resistivity as $\rho_{\rm Hall} = R_0 H + R_{\rm H} M$, 
with $R_0$ and $R_{\rm H}$ the ordinary and anomalous Hall 
coefficients, we find a  weak ordinary Hall resistance in 
FePd (Fig.~\ref{AHEmeasured}(a)), noticeable in the shallow 
slope of the saturation regions of the Hall loop. The ordinary
Hall coefficient $R_\mathrm{0}$ in the FePt sample is at
least an order of magnitude lower than in FePd at all temperatures,
and can safely be considered negligible in comparison to
the anomalous Hall signal itself, in this material being an 
order of magnitude larger than in FePd (see also Table~\ref{tabD}). 
The evolution of both the anomalous Hall resistivity $\rho_{\mathrm{H}}$
and the longitudinal resistivity $\rho_{\mathrm{0}}$ with temperature
was determined by extrapolating from the magnetically saturated state
back to zero field for each value of $T$. The negative sign of $R_0$ is
the same for both materials, as is the positive sign of the AHE
coefficient $R_{\rm H}$. Negative ordinary Hall coefficients 
correspond to electron-like transport. The ordinary and anomalous 
Hall coefficients are shown for two different temperatures in 
Table~\ref{tabD}.
\begin{figure}
\hspace{0.0cm}\includegraphics[width=7.4cm]{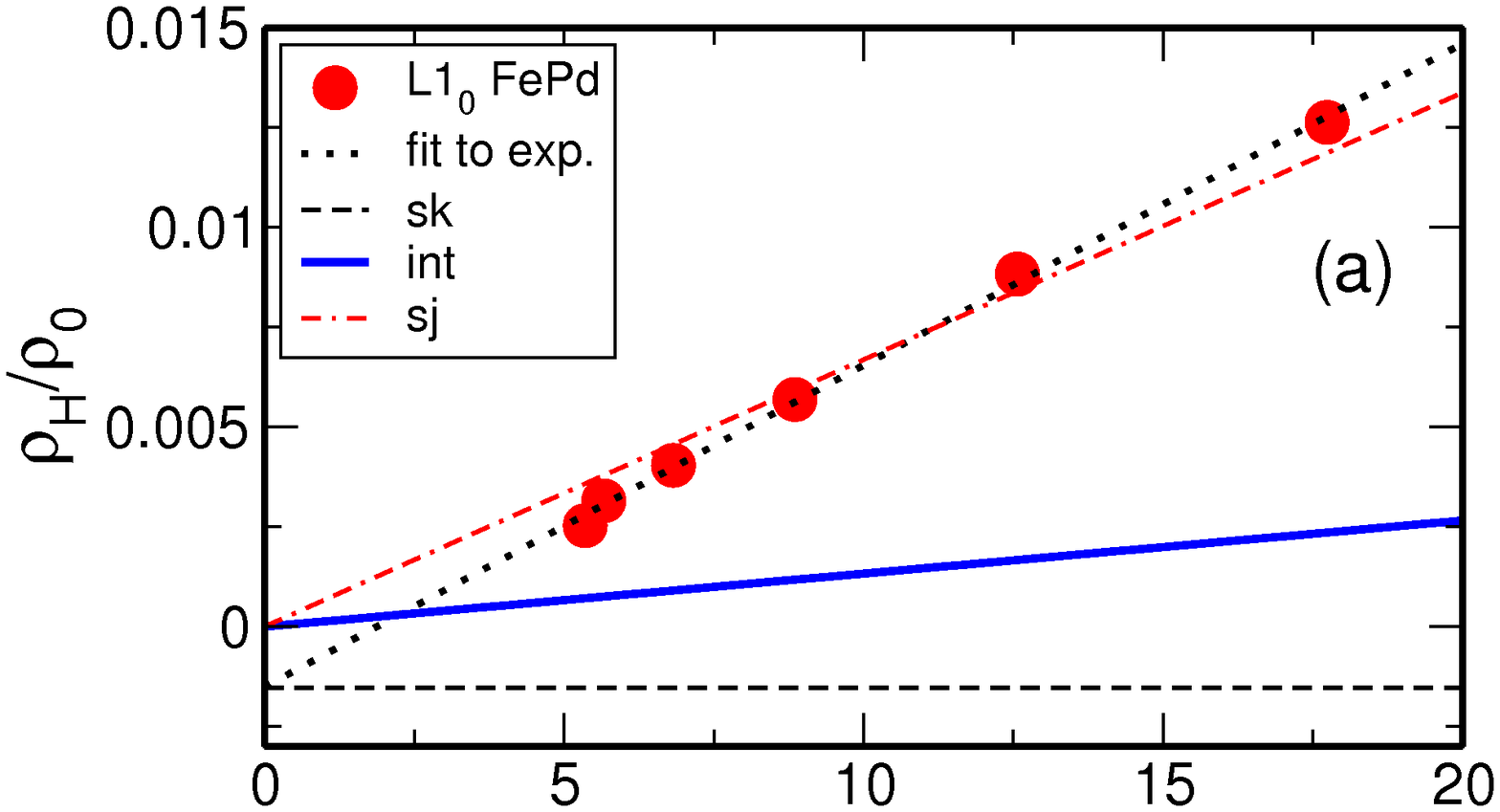}\vspace{0.2cm}
\includegraphics[width=7.1cm]{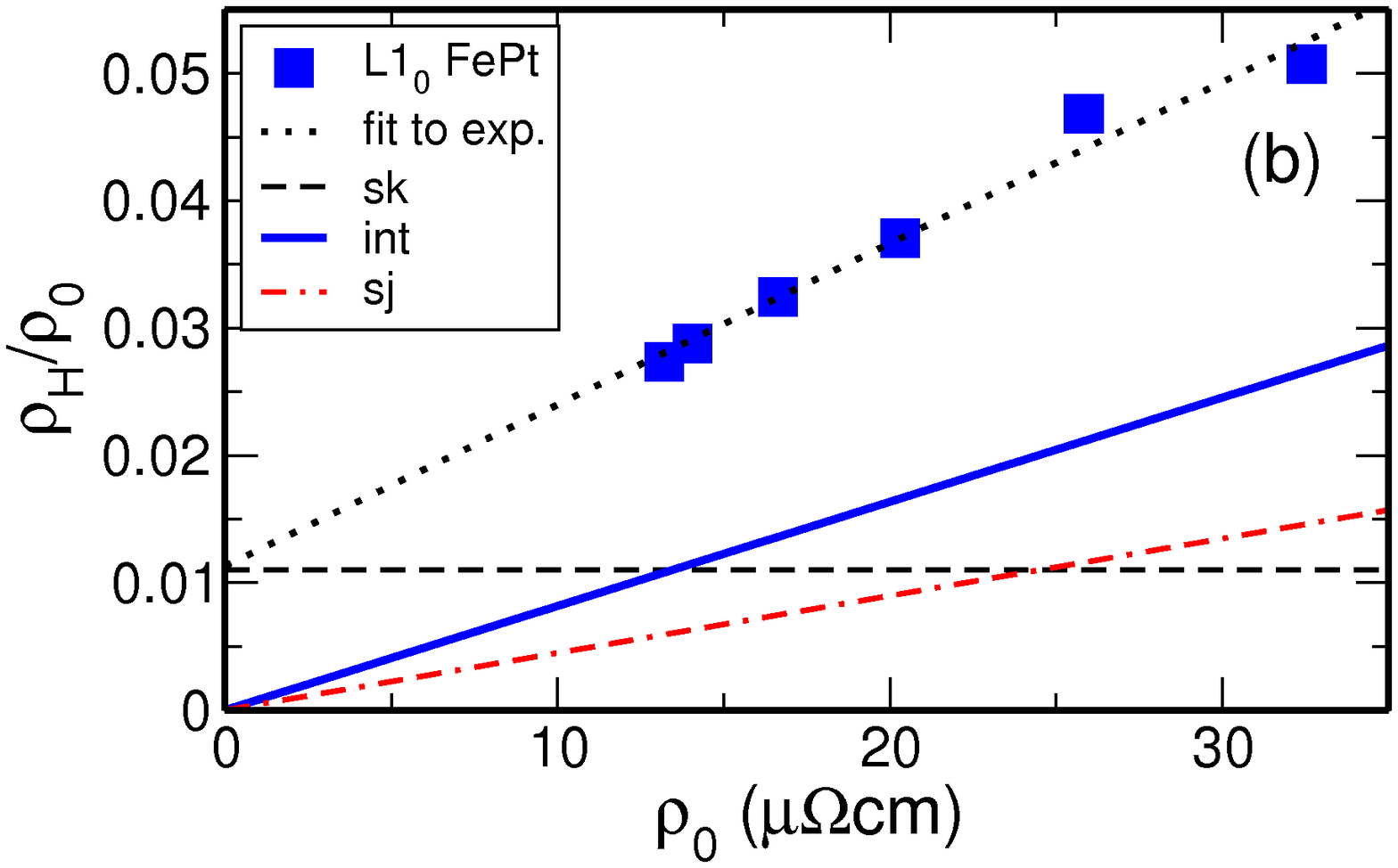}
\caption{(Color online) Resistivity ratio of measured transverse
and longitudinal resistivity $\rho_{\mathrm{H}}/ \rho_{\mathrm{0}}$
vs. longitudinal resistivity $\rho_{\mathrm{0}}$ for (a) \llo -FePd (circles)
and (b) \llo -FePt (squares). The thin black dotted line is a linear
$\Phi_{\mathrm{Sk}}+\kappa\rho_0$ fit to the measured ratio yielding 
$\Phi_{\mathrm{Sk}}$ (dashed line) and $\kappa$. 
The intrinsic contribution $\kappa^{\rm int}\rho_0$ (solid thick line) 
calculated from first principles as well as the side-jump contribution
$(\kappa-\kappa^{\rm int})\rho_0=\kappa^{\rm sj}\rho_0$ (dot-dashed line) 
yield a linear dependence in $\rho_{\mathrm{H}}/ \rho_{\mathrm{0}}$ vs. 
$\rho_{\mathrm{0}}$ (see also Table II). The total 
$\rho_{\mathrm{H}}/ \rho_{\mathrm{0}}$ signal is a sum of skew-scattering 
$\Phi_{\mathrm{Sk}}$, intrinsic and side-jump contributions.
\label{HallvsT}}
\end{figure}

In order to distinguish the various contributions to the experimental
AHE transport data we relate $\rho_{\mathrm{H}}$ to the longitudinal
resistivity $\rho_0$ employing a 
$\rho_{\mathrm{H}}=\Phi_{\mathrm{Sk}}\rho_{\mathrm{0}}+\kappa\rho_{\mathrm{0}}^2$
fit to experimental data for the two compounds~\cite{nagaosa2006}.
We assume that $\Phi_{\mathrm{Sk}}$ and $\kappa$ are constant and do not depend on
temperature $T$ via temperature-dependent magnetization~\cite{Weiterling},
as the measurements are performed at $T$ well below $T_C$.
According to Eq.~(\ref{eq1}), parameter $\kappa$ stands for the 
anomalous Hall conductivity and incorporates the intrinsic and side-jump
channels for the AHE, while $\Phi_{\mathrm{Sk}}$ is the skew-scattering
angle~\cite{footnote3}. In Fig.~3 we present the experimental $\rho_{\rm H}/\rho_0$ ratio 
plotted versus $\rho_0$, and its linear $\Phi_{\mathrm{Sk}}+\kappa\rho_0$ 
fit for FePd and FePt. This fit yields values for $\kappa$ in FePd and 
FePt of 806~S/cm~and 1267~S/cm, respectively. The extracted values of 
the skew-scattering angle $\Phi_{\mathrm{Sk}}$ for the two compounds,
given as an offset along the $y$-axis in Fig.~3, are opposite in sign, 
with $\Phi_{\mathrm{Sk}}$ of $-1.5$~mrad in FePd much smaller than that 
of $+11.0$~mrad in FePt (see Table~II). However, in our AHE experiments 
the $\rho_H$ resistivity is measured {\it in summa},~i.e.~its value remains
positive for the whole range of measured $\rho_0$.  

\begin{figure}[t!]
\hspace{-0.5cm}\includegraphics[width=6.0cm]{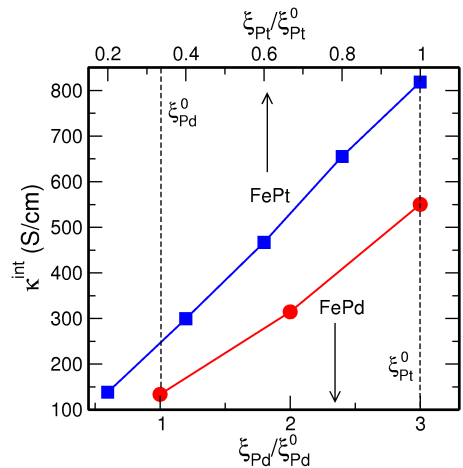}
\caption{(Color online)
Dependence of the {\it ab initio} calculated intrinsic anomalous Hall 
conductivity $\kappa^{\rm int}$ on the strength of the spin-orbit 
interaction inside the Pd atoms in $L1_0$-FePd, $\xi_{\rm Pd}$ (circles) 
and Pt atoms in $L1_0$-FePt, $\xi_{\rm Pt}$ (squares), with respect to 
their unscaled values $\xi_{\rm Pd}^0=0.19$~eV and $\xi_{\rm Pt}^0=0.54$~eV 
(dashed lines).}
\label{skewscangleFePdPt}
\end{figure}

In order to assess the intrinsic part of the AHE, we performed {\it ab 
initio} calculations for both materials, \llo -FePd and \llo -FePt,
in both cases for the perfectly ordered ($S_{\mathrm{Order}}$=1) systems, 
using the full-potential linearized augmented plane-wave (FLAPW) method 
as implemented in the J\"ulich density functional theory code {\tt
FLEUR}~\cite{fleur} and using the Wannier interpolation procedure of 
Wang {\it et~al.}~\cite{wang2006} for the Brillouin zone integration of the Berry 
curvature~\cite{footnote2}.  Our calculated values of the intrinsic 
anomalous Hall conductivity, $\kappa^{\rm int}$, are given in 
Table~II and constitute 818~S/cm~for FePt and 133~S/cm~for FePd. 
While in FePt the value of $\kappa^{\rm int}$ accounts for 65\% of 
experimentally measured $\kappa$, in FePd the intrinsic contribution 
constitutes only 15\% of the experimentally measured conductivity. 
This strongly suggests that if in FePt the intrinsic mechanism is 
the dominant source of $\kappa$, in isoelectronic FePd it is 
extrinsic side-jump scattering which provides a major contribution 
to it. Estimating the side-jump contribution to the conductivity 
$\kappa^{\rm sj}$ as the difference between the measured $\kappa$ 
and the calculated $\kappa^{\rm int}$, we observe a clear crossover 
between the two contributions to the conductivity $\kappa$ with 
$\kappa^{\rm int}({\rm FePt})>\kappa^{\rm sj}({\rm FePt})$ and 
$\kappa^{\rm int}({\rm FePd})\ll\kappa^{\rm sj}({\rm FePd})$, Table II.  
Overall, by comparing the intrinsic, side-jump and skew-scattering
contributions to the total measured $\rho_{\rm H}/\rho_0$ ratio,
presented in Fig.~3, we observe that while in FePd the measured
$\rho_{\rm H}$ resistivity is mainly due to side-jump, the AHE 
signal in FePt samples presents a competition between dominant 
intrinsic and extrinsic channels of the AHE.
\begin{table}
\begin{ruledtabular}
\caption{Coefficients of the linear $\Phi_{\mathrm{Sk}}+\kappa\rho_0$ 
fit to the experimental $\rho_H/\rho_0$ ratio for FePd and FePt (Fig.~2), 
the skew scattering angle $\Phi_{\mathrm{Sk}}$ and conductivity $\kappa$. 
$\kappa^{\rm int}$ stands for the calculated from {\it ab initio} value 
of the intrinsic conductivity, while the side-jump conductivity 
$\kappa^{\rm sj}$ is obtained as $\kappa-\kappa^{\rm int}$.}\label{tab2}
\begin{tabular}{lllll}
     & $\Phi_{\mathrm{Sk}}$ & $\kappa$ & $\kappa^{\rm int}$ 
     & $\kappa^{\rm sj}$  \\
 & mrad & S/cm & S/cm & S/cm \\
\hline
  FePd & $-1.5 \pm 0.2$ & $806\pm18$   & 133 & 669 \\
  FePt & $+11.0 \pm 2$  & $1267\pm101$ & 818 & 449 \\
\end{tabular}
\end{ruledtabular}
\end{table}

The crossover between the intrinsic-dominated AHE in FePt and the
side-jump domaninated effect in FePd is caused mainly by the
different spin-orbit coupling strengths in the two materials. 
We show this by artificially scaling the spin-orbit strength $\xi$ 
inside the Pt and Pd atoms, which is determined as an average of 
SOI matrix elements calculated on corresponding $d$-orbitals.
Figure~4 shows the calculated intrinsic anomalous Hall conductivity 
(AHC) of FePd as the SOI strength $\xi_{\rm Pd}$ inside the Pd 
atoms is artificially increased three-fold until it reaches the 
actual value in the Pt atoms. It is seen that the intrinsic AHC in 
FePd increases from 133~S/cm to 550~S/cm, which is comparable to the 
value of 818~S/cm calculated in FePt. Conversely, upon decreasing 
the SOI strength in the Pt atoms by a factor of three, the AHC of 
FePt decreases to about 250~S/cm. The remaining differences between 
the intrinsic AHC in FePt and FePd can be accounted for by their 
different lattice constants, orbitals spreads, and Stoner parameters.

On the other hand, according to Table~II, while $\kappa^{\rm int}$ increases 
the side-jump conductivity $\kappa^{\rm sj}$ decreases moderately
with increasing SOI strength. Such behavior can be seen from model 
calculations for a 2D model Hamiltonian, namely, massive Dirac Hamiltonian with
randomly distributed weak $\delta$-function-like spin-independent
impurities~\cite{Sinitsyn2007}. In this case the ratio of the intrinsic
and side-jump conductivities can be deduced analytically and constitutes 
$\kappa^{\rm int}/\kappa^{\rm sj}=\xi^2+1/4$, where $\xi=\Delta/(vk_F)$
is the scaled SOI strength $\Delta$, $v$ is the parameter of the model, and $k_F$ 
is the Fermi wave vector determined by the band filling~\cite{Sinitsyn2007}. 
It is clear that, while for large 
SOI the intrinsic $\kappa$ dominates over $\kappa^{\rm sj}$, for 
comparatively small $\xi$ the side-jump can be several times larger 
than $\kappa^{\rm int}$ within this model~\cite{footnote}. In our work, 
we provide the first experimental verification of this phenomenon in a 
bulk material with a complex electronic structure. Moreover, we suggest 
that this phenomenon may be employed to adjust the ratio of intrinsic 
and extrinsic contributions to the AHE 
with the prospect of engineering materials with desired AHE properties. 
For example maximizing AHE in the FePt$_x$Pd$_{1-x}$ L1$_0$
alloys~\cite{JAP02} by adjusting the ratio between side-jump and intrinsic 
AHE and tuning the skew-scattering by impurity concentration at the same 
time may lead to novel devices.

%For example maximizing AHE in the FePt$_x$Pd$_{1-x}$ L1$_0$ 
%alloys~\cite{JAP02} by adjusting extrinsic skew and side-jump 
%scattering and intrinsic AHE at the same time may lead to novel 
%devices. 

\begin{acknowledgments}
The authors thank G. Obermeier, J. Cunningham, M.C. Hickey, A. Blackburn 
for fruitful discussions and P. Cale, G. Butterworth and T. Haynes for 
technical assistance. Financial support by the UK EPSRC (Spin@RT), by the
BMBF (05KS4UK1/4) and the DFG (project MO~1731/1-1) 
is gratefully acknowledged.
\end{acknowledgments}

\bibliographystyle{apsrev_three}

%\bibliography{FePdPtanohall}

\end{document}